\documentclass[pre,twocolumn,showpacs,preprintnumbers,amsmath,amssymb]{revtex4}
\usepackage{graphicx}
\usepackage{dcolumn}
\usepackage{bm}

\def \half{ \frac {1}{2} }
\def \be{\begin{equation}}
\def \ee{\end{equation}}
\def \bea{\begin{eqnarray}}
\def \eea{\end{eqnarray}}
\def \ba{\begin{array}}
\def \ea{\end{array}}
\def \non{\nonumber}
\begin{document}


\title{Energy distribution and effective temperatures in a driven dissipative model}
\author{Yair Shokef}
\thanks{Formerly: Srebro}
\author{Dov Levine}
\affiliation{Department of Physics, Technion, Haifa 32000, Israel}
\date{\today}

\begin{abstract}

We investigate non-equilibrium behavior of driven dissipative systems, using the model presented in [Phys. Rev. Lett. {\bf 93}, 240601 (2004)]. We solve the non-Boltzmann steady state energy distribution and the temporal evolution to it, and find its high energy tail to behave exponentially. We demonstrate that various measures of effective temperatures generally differ. We discuss infinite hierarchies of effective temperatures defined from moments of the non-exponential energy distribution, and relate them to the ``configurational temperature'', measured directly from instantaneous particle locations without any kinetic information. We calculate the ``granular temperature'', characterizing the average energy in the system, two different ``fluctuation temperatures'', scaling fluctuation-dissipation relations, and the ``entropic temperature'', defined from differentiating the entropy with respect to energy. 

\end{abstract}


\maketitle

\section{Introduction}

Systems of many particles interacting dissipatively are far from thermodynamic equilibrium, and a general theoretical description of their statistical mechanics is lacking, in contrast to systems in equilibrium, for which there is a well established theory. In this Paper we use exact solutions of a simple stochastic model in order to explore aspects of dissipative systems. In particular, we are interested in steady states: their non-Boltzmann energy distribution, the way in which a system arrives at its steady state, and various proposed {\it definitions} of temperature. Previous theoretical research on energy distributions and effective temperatures in various driven dissipative systems (granular materials \cite{noije_1998,bennaim_krapivsky_2000,moon_2001,moon_2004,zon_2004,zon_2005,puglisi_2002_fd,barrat_2002_fd,garzo_2004}, foams \cite{langer_2000,t_fd_foam,ohern_2004}, and glasses \cite{barrat_2000,barrat_2001,berthier_2002}) has been mostly numerical or approximate. Therefore the resulting distributions and temperatures may agree with an effective equilibrium behavior due to numerical error.

In systems comprised of macroscopic particles, energy is dissipated via interactions, being transferred from macroscopic degrees of freedom (such as motion of particles) into microscopic degrees of freedom (heat), and can not be transformed back. Continuous driving is needed in order to maintain such a system in a dynamic state. One way to model this driving is by holding the system in contact with a bath, or large energy reservoir. As in \cite{biben_2002}, we concentrate on driving mechanisms where this bath is in equilibrium at some temperature $T_B$ (this driving bath is not necessarily in thermal equilibrium). A non-dissipative system driven by a thermal bath would reach thermodynamic equilibrium with it, where the energy distribution is given by the exponential Boltzmann distribution, and the system temperature is equal to the bath temperature. 

The temperature of an equilibrium system is manifested in various measurements that can be performed on it. A non-equilibrium system does not, a-priori, have a unique well-defined temperature, and each such measurement inspires the definition of a corresponding effective temperature. For example, the {\it entropic temperature} $T_S$ is the inverse of the derivative of entropy with respect to energy \cite{barrat_2000,coniglio_2001,makse_2002,t_fd_foam}, in analogy with the definition of temperature in statistical mechanics. The principle of energy equipartition in thermal equilibrium motivates defining the {\it granular temperature} $T_G$ as the average energy per degree of freedom \cite{tg}; The equilibrium fluctuation-dissipation theorem suggests defining the {\it fluctuation temperature} $T_{F}$ as the ratio of fluctuation to response \cite{hohenberg_1989,cugliandolo_1997}; Recent results expressing the equilibrium temperature from ensemble averages of particle locations (without any kinetic information) \cite{rugh_1997,jepps_2000} lead to the definition of a {\it configurational temperature} $T_C$ as another measure of effective temperature for non-equilibrium systems \cite{han_grier_2004,powels_2005}. All these definitions yield the same value in equilibrium.

In steady states of driven dissipative systems all effective temperatures are generally much smaller than $T_B$ ($T_B$ should not be confused with the actual temperature of the environment which is typically much lower), and, unlike thermal equilibrium, their values depend on the details of the coupling with the bath. Although these systems are far from equilibrium, and their energy distributions differ significantly from the Boltzmann distribution, there is evidence for coincidence of different effective temperatures. Different $T_{F}$s of the same system, obtained from correlations and response of different variables, have been found in numerical experiments to coincide in glasses \cite{barrat_2000,barrat_2001,berthier_2002}, and to coincide with $T_G$ in granular gases \cite{puglisi_2002_fd,barrat_2002_fd} and with $T_S$ in sheared foam \cite{t_fd_foam}.

In this Paper we study an exactly solvable dissipative model, in which interactions occur randomly and redistribute energy stochastically between the interacting particles. In \cite{srebro_levine} we first introduced the model, discussed its similarities to granular gases, calculated all moments of its steady-state energy distribution, and showed that $T_{F}$ differs from $T_G$. Here we present these results in more detail and provide further results on the model: We define the model in Sec. \ref{sec:definition} and investigate its energy distribution in Sec. \ref{sec:model_sol}. We use the energy scales defined by high moments of the energy distribution to show that the high energy tail is exponential with a decay rate corresponding to the bath temperature $T_B$, and solve the temporal evolution to the steady state. Section \ref{sec:teffs} deals with various effective temperatures in our model. We show that spontaneous fluctuations of different quantities are scaled by different fluctuation temperatures. We demonstrate that for systems with a single energy scale and with smooth energy distributions, the entropic temperature coincides with the granular temperature, and then investigate cases where they differ. We relate the infinite hierarchies of different effective temperatures defined from the energy distribution to generalizations of the configurational temperature. All results are calculated exactly in the context of our model, and in Sec. \ref{sec:discussion} we discuss the generality of these results to other driven dissipative systems.

\section{Model Definition}
\label{sec:definition}

At its most basic, a dissipative system consists of a set of degrees of freedom, or modes of excitation, which interact among themselves and with the external environment. When two modes interact with each other, there is energy exchange, with some of the energy being lost to the environment. External driving may be thought of as the injection of energy into the system's modes from the environment. Our model is constructed as a minimal model including these essential features. When two particles (we shall refer to the modes as particles for simplicity) interact they lose some of their energy and exchange what remains, and when a particle interacts with the environment it is more likely to gain energy rather than to lose.

\begin{figure}[t]
\includegraphics[width=8cm]{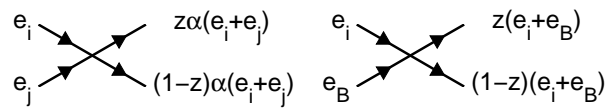}
\caption{\label{fig:model_diag} The possible interactions in our model: dissipative two particle interaction (left) and conservative system-bath interaction (right).}
\end{figure}

Our model consists of $N$ particles with energies $\{ e_i \}$, with a constant interaction rate between any two particles in the system. In every interaction two particles from the system are chosen at random and their energies are summed. In the case of conservative dynamics (analogous to elastic collisions) this total energy is repartitioned randomly between the two interacting particles (as in \cite{ulam_1980}). For dissipative dynamics with inelastic collisions, only a fraction $0 \leq \alpha < 1$ of the total energy is repartitioned between the particles and the rest is dissipated out of the system. Thus, $\alpha$ is analogous to a restitution coefficient. The system is coupled to a heat bath so that it may be maintained in a nontrivial steady state. The interactions are shown diagrammatically in Fig. \ref{fig:model_diag}, and described in further detail below.

For conservative dynamics ($\alpha=1$) an isolated system (i.e., not in contact with the driving bath) reaches thermodynamic equilibrium with the exponential Boltzmann distribution for each particle's energy, $p(e)=T^{-1} \exp(-e/T)$, where the temperature equals the average energy per particle $T = \langle e \rangle$ (we measure temperature in units of energy and set Boltzmann's constant to one). Dissipative dynamics ($0 \leq \alpha<1$) cause energy to decay, therefore we drive the system by attaching it to a heat bath, constructed as an infinitely large system of particles obeying the conservative dynamics described above, kept in equilibrium at a temperature $T_{B}$. The coupling of the dissipative system to the bath is through conservative interactions between a particle chosen at random from the system and a particle chosen at random from the bath (the system-bath interactions are taken as conservative for simplicity, however dissipative interactions may as well be considered, yielding qualitatively similar results). This contact is characterized by a coupling strength, $0 < f \leq 1$, defined as the fraction of every particle's interactions with the bath out of all its interactions. Unlike thermodynamic equilibrium, the dissipative system's steady state depends on the bath through both $T_B$ and $f$ (see \cite{zon_2004} which emphasizes the importance of coupling details).

The stochastic evolution of the energy of particle $i$ during an infinitesimal time step $dt$ is hence given by
\bea e_i(t+dt)= \left\{
\begin{array}{cc}
  \underline{Value}: & \underline{Probability}: \\
  e_i(t) & 1- \Gamma dt \\
  z \alpha \left[ e_i(t)+e_j(t) \right]  & (1-f) \Gamma dt \\
  z \left[ e_i(t)+e_B \right]  & f \Gamma dt
\end{array} \right.\label{eq:dyn_rule},
\eea
where $\Gamma$ is the interaction rate per particle per unit time, $j \in \{ 1 , ..., N \}$ ($j \ne i$) is the index of the particle with which particle $i$ may interact (chosen randomly at every interaction), $z \in [0,1]$ is the fraction of repartitioned energy given to particle $i$ in the interaction (chosen randomly with a uniform distribution at every interaction), and $e_{B}$ is the energy of the bath particle with which particle $i$ may interact, which at every interaction is chosen randomly from the equilibrium distribution in the bath: $p_B(e_B) = T_B^{-1} \exp(-e_B/T_B)$.

The simplicity of our model derives from the fact that every particle in it is described only by its energy, as opposed, for example, to the $2d$ degrees of freedom per particle in a $d$-dimensional frictionless hard sphere gas. By eliminating the momentum and spatial variables and using only the energy, we replace the vectorial collisions between particles by scalar interactions. Furthermore, since any two particles may interact, there are no spatial correlations.

\section{Energy Distribution}
\label{sec:model_sol}

In this section we investigate our model's single-particle energy distribution $p(e)$. We calculate from Eq. (\ref{eq:dyn_rule}) the temporal evolution of any moment of $p(e)$. This is used to obtain the temporal evolution of $p(e)$, as well as its form in steady state. We use the energy moments to define two hierarchies of energy scales, whose asymptotic behavior is then used to characterize the high energy tail of $p(e)$. In particular, we shall show that although $p(e)$ differs from the Boltzmann distribution, its high energy tail is exponential.

\subsection{Average Energy}

\begin{figure}[t]
\includegraphics[width=7cm]{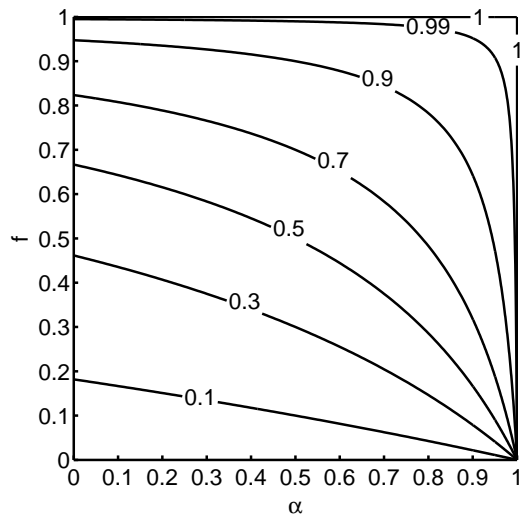}
\caption{\label{fig:tg_contours} Contours of the ratio $\langle e \rangle/T_B$ between the average energy per particle in the system and the bath temperature vs. the restitution coefficient $\alpha$ and the coupling strength $f$ to the bath, as given by Eq. (\ref{eq:tg_steady}).}
\end{figure}

The most direct way to characterize the system's state is by the average energy per degree of freedom. It is solved by averaging Eq. (\ref{eq:dyn_rule}) over the stochasticity in the dynamics and over all particles:
\bea \langle e (t+dt) \rangle &=& (1- \Gamma dt) \langle e(t) \rangle + (1-f) \Gamma dt \alpha \langle e(t) \rangle \non \\ &+& f \Gamma dt \half \left[ \langle e(t) \rangle +T_B \right] \label{eq:1st_moment}.
\eea
Hence $\langle e(t) \rangle$ satisfies the differential equation
\bea \frac{2}{\Gamma} \frac{d\langle e(t) \rangle}{dt} =  - A_1 \langle e(t) \rangle + f T_B \label{eq:tg_diff_eq},\eea
with $A_1(\alpha,f) \equiv 1 + (1-f) (1-2 \alpha) > 0$. This has the steady state solution 
\be \langle e \rangle = \frac{f T_B}{A_1} = \frac{T_B} {2 \alpha -1 + 2 (1-\alpha)/f }
\label{eq:tg_steady}. \ee
In analogy with granular materials, this is denoted as the granular temperature $T_G \equiv \langle e \rangle$. It is plotted vs. $\alpha$ and $f$ in Fig. \ref{fig:tg_contours}.

It is interesting to note that despite the simplicity of our model, it captures, at least qualitatively, some aspects of actual driven dissipative systems \cite{rouyer_2000,feitosa_menon_2002}. For instance, the stationary value of $\langle e \rangle$ is always smaller than $T_B$ and depends not only on the dissipation through the restitution coefficient $\alpha$, but also on the details of the coupling to the bath through the coupling strength $f$. $\langle e \rangle$ coincides with $T_B$ only in the two non-dissipative limits: conservative interactions ($\alpha=1$) and strong coupling to the bath ($f=1$). 

\subsection{Energy Fluctuations}

We now consider the energy distribution around the average energy $\langle e \rangle$. This tests whether the system is equivalent to an equilibrium system at an effective temperature $T_G = \langle e \rangle$. That is, whether the effect of the dissipation is to modify the Boltzmann distribution only by changing its characteristic temperature from $T_B$ to $T_G$. However, this is not the case: we find that the energy distribution is clearly non-exponential, in qualitative agreement with realistic driven dissipative systems, such as granular gases \cite{noije_1998,rouyer_2000,moon_2001,moon_2004} and colloidal suspensions \cite{kob_1997,kasper_1998,marcus_1999,weeks_2000}.

The first moment of $p(e)$ is the average energy calculated above. Higher moments are obtained by taking the average of the $n$th power of Eq. (\ref{eq:dyn_rule}). This yields the following differential equation for $\langle e^n \rangle$ in terms of all lower moments and the moments of the energy distribution in the bath (for which $\langle e_B^m \rangle = m ! T_B^m$),
\bea  \frac{n+1}{\Gamma} \frac{d \langle e^n \rangle}{dt} = -A_n \langle e^n \rangle \non \\ + \sum^{n-1}_{m=1} { n \choose m } \langle e^m \rangle \left[ (1-f) \alpha ^n \langle e^{n-m} \rangle + f \langle e_B^{n-m} \rangle \right] \non \\ + f \langle e_B^n \rangle , \label{eq:n_mom_diff}
\eea
where 
\bea
A_n(\alpha,f) \equiv n + (1-f)(1-2 \alpha^n) \label{eq:An}. 
\eea
Any initial distribution will evolve with time to the steady state distribution given by 
\bea \langle e^n \rangle = \{ \sum^{n-1}_{m=1} {n \choose m} \langle e^m \rangle \left[ (1-f)\alpha ^n \langle e^{n-m} \rangle + f \langle e_B^{n-m} \rangle \right] \non \\ + f \langle e_B^n \rangle \} / A_n \label{eq:moments_recursion} .\eea
These expressions for all energy moments are exact arbitrarily far from equilibrium (for general values of $\alpha$ and $f$) and contain information about the entire energy distribution. In the equilibrium limits ($\alpha=1$ or $f=1$) Eq. (\ref{eq:moments_recursion}) yields the moments of the exponential Boltzmann distribution, for which $\langle e^n \rangle = n! T_B^n$. 

\begin{figure}[h]
\includegraphics[width=8cm]{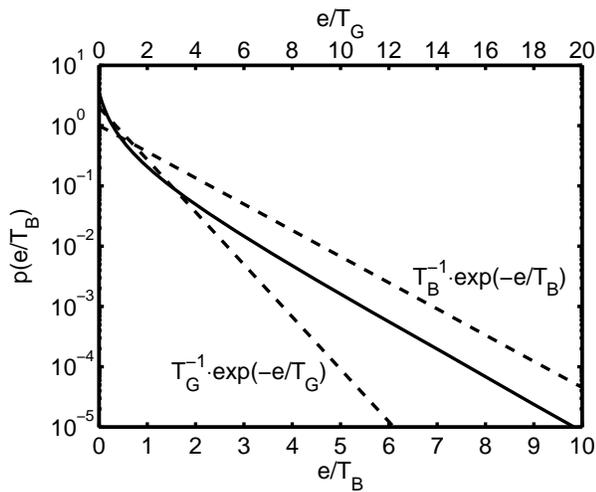}
\caption{\label{fig:ener_dist} Steady state energy distribution for restitution coefficient $\alpha=0.5$ and coupling strength $f=0.5$ (resulting in $T_G=T_B/2$) obtained in a numerical simulation of the model (solid line). Exponential distributions at temperatures $T_B$ and $T_G$ are given for reference (dashed lines).}
\end{figure}

The steady state energy distribution with moments given by Eq. (\ref{eq:moments_recursion}) is shown in Fig. \ref{fig:ener_dist}. At low energies (of the order of several times $T_G$) the distribution is roughly exponential with a decay rate corresponding to the average energy $T_G$. At intermediate energies the distribution seems to exhibit an overpopulated high energy tail, decaying slower than $\exp(-e/T_G)$. However, for higher energies ($e \gtrsim 10 T_G = 5 T_B$ for the parameters in Fig. \ref{fig:ener_dist}), the distribution exhibits an exponential decay of the form 
\bea
p(e) \sim \exp(-e/T_B). 
\eea
The overpopulation of the high energy tail is only with respect to scaling the energy with the average energy in the system and considering energies comparable to $\langle e \rangle$, as is customarily done in granular gases \cite{noije_1998,zon_2004,zon_2005,moon_2001,moon_2004,rouyer_2000}. Interactions with the bath dominate the high energy tail, since a particle is much more likely to arrive at such high energies due to a conservative interaction with the bath rather than due to a dissipative interaction within the system (and the fraction of very high energy particles in the bath is larger than in the system). In the next section, we use the asymptotic behavior of energy scales defined by the moments to show that the tail is indeed exponential.

\subsection{High Energy Tail}
\label{sec:tail}

\begin{figure}[b]
\includegraphics[width=8cm]{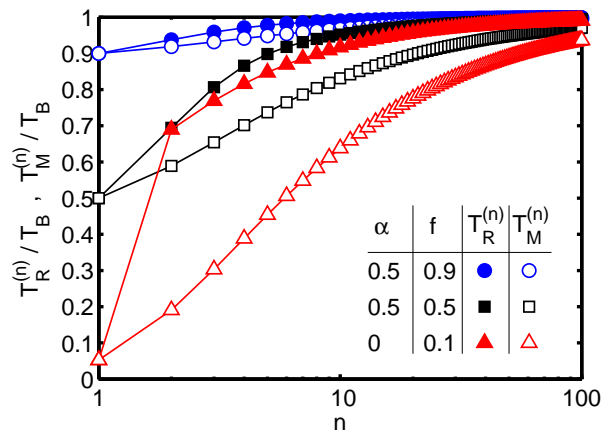}
\caption{\label{fig:pl_moments} (Color online) The hierarchies of effective temperatures $T_R^{(n)}$ and $T_M^{(n)}$ for several model parameters, obtained by substituting Eq. (\ref{eq:moments_recursion}) in Eq. (\ref{eq:def_tr_tm}).}
\end{figure}

The exponential energy distribution of a system in equilibrium contains a single energy scale - the temperature of that system. For a non-equilibrium system, with a non-exponential energy distribution, infinite hierarchies of effective temperatures may be defined from the energy distribution. We define two such hierarchies, $T_R^{(n)}$ from the ratios of succeeding energy moments, and $T_M^{(n)}$ by scaling the moments themselves:
\begin{subequations} \label{eq:def_tr_tm}
\bea
T_R^{(n)} \equiv \frac {\langle e^n \rangle} {n \langle e^{n-1} \rangle} , \label{eq:def_tr}
\eea
\bea
T_M^{(n)} \equiv \left( \frac {\langle e^n \rangle} {n!} \right)^{1/n} . \label{eq:def_tm}
\eea
\end{subequations}
Both reduce in equilibrium to the system's temperature for any $n$. Away from equilibrium they typically differ and depend on $n$ (see Fig. \ref{fig:pl_moments}), and for $n=1$ both reduce to the granular temperature $T_G \equiv \langle e \rangle$. 

The large $n$ behavior of $T_R^{(n)}$ and $T_M^{(n)}$ reflects the distribution's high energy tail. For an exponential tail $p(e) \sim \exp(-e/T_{eff})$, $T_R^{(\infty)}=T_M^{(\infty)}=T_{eff}$; For a stretched exponential tail, $p(e) \sim \exp(-c e^a)$, $T_R^{(\infty)}=T_M^{(\infty)}=0$ if $a>1$, and $T_R^{(\infty)}=T_M^{(\infty)}=\infty$ if $a<1$; For a power-law tail $T_M^{(\infty)}=\infty$ while $T_R^{(\infty)}$ is undefined. 

We now show that the energy moments in our model are consistent with an exponential high energy tail with a decay constant corresponding to the bath temperature $T_B$. Since $0 \leq \alpha \leq 1$ for $n \gg 1$, $A_n \approx n$ [see Eq. (\ref{eq:An})]. Furthermore, for $0 < f \leq 1$, since $0 \leq \alpha < 1$, we have $(1-f) \alpha ^n \langle e^{n-m} \rangle \ll f \langle e_B^{n-m} \rangle$, and Eq. (\ref{eq:moments_recursion}) reduces to
\bea 
\langle e^n \rangle \approx f (n-1)! \sum^{n-1}_{m=0} \frac{\langle e^m \rangle}{m!} T_B^{n-m} \label{eq:moments_recursion_aprox} , 
\eea
where the term $f \langle e^n_B \rangle$ in Eq. (\ref{eq:moments_recursion}) has been incorporated as the $m=0$ term in the summation in Eq. (\ref{eq:moments_recursion_aprox}). The solution for large $n$ may be approximated by taking the continuum limit, where the sum in Eq. (\ref{eq:moments_recursion_aprox}) transform to an integral, and the resulting equation may be solved to yield
\bea
\langle e^n \rangle \approx C n! T_B^n n^{f-1} 
\eea
with $C$ a dimensionless constant independent of $n$. Therefore, in the large $n$ limit $T_R^{(n)}$ and $T_M^{(n)}$ both converge to $T_B$, 
\begin{subequations}
\bea
T_R^{(n)} = T_B \left( \frac{n}{n-1} \right)^{f-1} \rightarrow T_B 
\eea
\bea
T_M^{(n)} = T_B \left( C n^{f-1} \right)^{1/n} \rightarrow T_B 
\eea
\end{subequations}
as can also be seen in Fig. \ref{fig:pl_moments}. This supports the observation that the high energy tail behaves as $\exp(-e/T_B)$. 

It is intriguing to speculate on the generality of this result, that for very high energies the distribution behaves as an equilibrium distribution with a temperature equal to the bath temperature $T_B$ (see also \cite{biben_2002} where similar results have been found for a granular gas driven by an ideal gas heat bath).

\subsection{Approach to Steady State}

We solve Eq. (\ref{eq:n_mom_diff}) recursively with $n$, and find that the time-dependent solution is of the form 
\bea \langle e^n(t) \rangle = \langle e^n \rangle + \sum_{m=1}^n C_{n,m} \exp \left( -\frac{A_m \Gamma}{m+1} t \right), \eea
where $\langle e^n \rangle$ are the steady state moments given by Eq. (\ref{eq:moments_recursion}), and $\{ C_{n,m} \}$ are constants depending on the initial distribution. $A_m$ is discrete and increases monotonically with $m$, therefore the slowest exponential decay with time, $\exp \left( - A_1 \Gamma t / 2 \right)$, dominates the long time behavior of all moments. When scaling high moments to units of energy, one has 
\bea
T_M^{(n)} \sim \langle e^n(t) \rangle ^{1/n} \sim \exp \left( - \frac{A_1 \Gamma}{2n} t \right). 
\eea
Thus the high energy components of the distribution approach their steady values slower than the low energy ones. It is interesting to note that this is similar to the analysis of \cite{krook_wu} for the approach to equilibrium in plasmas. 

\begin{figure}[h]
\includegraphics[width=8cm]{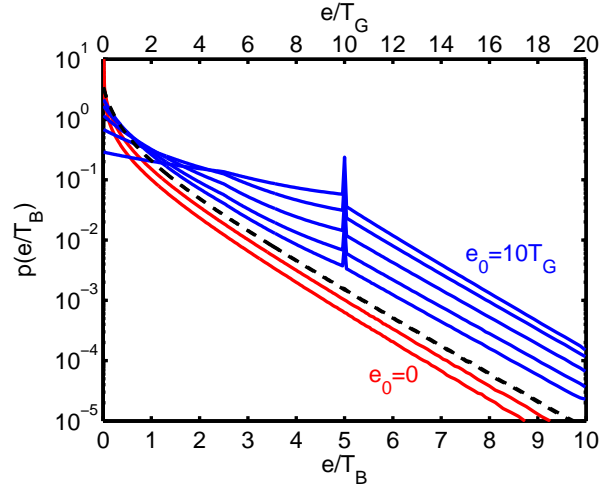}
\caption{\label{fig:transient} (Color online) Temporal evolution to the steady state energy distribution (dashed line) for restitution coefficient $\alpha=0.5$ and coupling strength $f=0.5$ obtained in a numerical simulation of the model, starting with all particles having energy $e_0=0$ (ascending lines after $N$ and $2N$ interactions in the system), or $e_0=5T_B$ (descending lines after $N$, $2N$, $3N$, $4N$, and $5N$ interactions).}
\end{figure}

Figure \ref{fig:transient} provides the temporal evolution of the energy distribution starting from two different initial conditions. In both cases the system started with all particles having the same energy, $e_0=0$ in one case, and $e_0=5T_B$ in the other, and reaches its stationary distribution within several interactions per particle. For $e_0=0$ this approach is uniform and faster. For $e_0=5T_B$ the system needs a slightly longer time until it arrives to the steady state, and the distribution behaves differently on both sides of $e_0$. For $e>e_0$ single particle interactions with the bath dominate, and the system exhibits the bath dominated exponential tail. For $e<e_0$, on the other hand, the two-body dissipative interactions require multiple collisions in order to change the shape of the distribution continuously from the initial delta function to the smooth steady state distribution.

\section{Effective Temperatures}
\label{sec:teffs}

Various definitions of effective temperatures are used to characterize systems far from equilibrium. One of the important questions to be answered in this context is to what extent do effective temperatures defined by different measurements on a single system yield the same numerical value. In Sec. \ref{sec:model_sol} we first defined the granular temperature as the average energy per degree of freedom $T_G \equiv \langle e \rangle$, and then used the energy moments to define two generalizations of $T_G$ to the hierarchies $T_R^{(n)}$ and $T_M^{(n)}$ of effective temperatures [Eq. (\ref{eq:def_tr_tm})]. For a non-exponential energy distribution these generally differ and moreover depend on $n$ (see Fig. \ref{fig:pl_moments}). In this section we investigate the interrelations between three additional definitions of effective temperatures - the fluctuation temperature $T_{F}$, the entropic temperature $T_S$, and the configurational temperature $T_C$.

\subsection{Fluctuation Temperature}
\label{sec:tf}

In equilibrium the fluctuation-dissipation theorem may be used to deduce a system's temperature from the spontaneous fluctuations of any of its physical quantities. The theorem assures that such a fluctuation is equal to the temperature multiplied by the corresponding response function or susceptibility. As such, fluctuations and susceptibilities of various quantities in non-equilibrium systems may be similarly used to define effective temperatures. Here we calculate two such fluctuation temperatures and show that they generally differ one from the other as well as from the granular temperature $T_G$ and bath temperature $T_B$.

To measure fluctuation-dissipation relations, we add degrees of freedom $\{ x_i \}$ to our model, whose correlations may be measured, and upon which a response measurement may be performed. To this end we couple the $\{ x_i \}$ to an external field $F_i$. That is, we modify the total energy of particle $i$ to $e_i - x_i F_i$, and refer to $\{ e_i \}$ as ``kinetic'' energies (or the {\it system}) and to $\{ -x_i F_i(t) \}$ as ``internal'' energies  (or the {\it probe}). (This probe is added only for measuring fluctuation-dissipation relations as described in this section; all other sections of the Paper deal with the model defined in Sec. \ref{sec:definition} without the ``internal'' degrees of freedom.)

\begin{figure}[b]
\includegraphics[width=5cm]{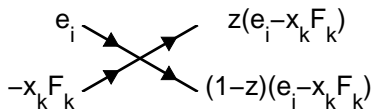}
\caption{\label{fig:model_diag_xF} The interactions between the system $\{ e_i \}$ and the probe $\{ x_i \}$, introduced in addition to those given in Fig. \ref{fig:model_diag}.}
\end{figure}

We assume the driven dissipative dynamics defined in Sec. \ref{sec:definition} for the system together with non-dissipative exchange of ``kinetic'' and ``internal'' energy, as described by Fig. \ref{fig:model_diag_xF}. This yields the following stochastic equations of motion:
\begin{subequations}\label{eq:dyn_rule_fd}
\bea e_i(t+dt)= \left\{
\begin{array}{cc}
  \underline{Value}: & \underline{Probability}: \\
  e_i(t) & 1- \Gamma dt \\
  z \alpha \left[ e_i(t)+e_j(t) \right]  & (1-f) \Gamma dt \\
  z \left[ e_i(t)+e_{B} \right] & f (1-h) \Gamma dt \\
  z \left[ e_i(t)-x_k(t) F_k(t) \right] & f h \Gamma dt
\end{array} \right.\label{eq:dyn_rule_fd_e} 
\eea
\bea x_i(t+dt)= \left\{
\begin{array}{cc}
  \underline{Value}: & \underline{Probability}: \\
  x_i(t) & 1- f h \Gamma dt \\
  z \left[ x_i(t)-e_k(t) / F_i(t) \right]  & f h \Gamma dt
\end{array} \right.\label{eq:dyn_rule_fd_x} 
\eea
\end{subequations}
where $h$ is a parameter introduced to describe the coupling strength between the system and the probe: $0<h<1$ is the fraction of the system's interactions with the probe out of all its non-dissipative interactions (with the bath and with the probe), and $k \in \{ 1 , ..., N \}$ is the index of the particle with which particle $i$ may interact in a ``kinetic''-``internal'' interaction [this is denoted differently from the index $j$ of the second particle in a ``kinetic''-``kinetic'' interaction, since $e_i$ may not interact with $e_i$ (thus $j \ne i$), while $e_i$ and $x_i$ may interact ($k$ may take the value $i$)].

We examine two measurements testing the relation between steady state fluctuation and response. First, we consider the fluctuation $\langle \Delta x^2 \rangle \equiv \langle x_i^2 \rangle - \langle x_i \rangle ^2$ of a single particle's $x_i$, and its response $r$ with respect to a change in $F_i$, $r \equiv \partial \langle x_i \rangle / \partial F_i$. Second, we define the total system's $X \equiv \sum_{i=1}^N x_i$ and consider the relation between its fluctuation $\langle \Delta X^2 \rangle \equiv \langle X^2 \rangle - \langle X \rangle ^2$, and its response $R$ with respect to a change in the uniform field $F$, $R \equiv \partial \langle X \rangle / \partial F$. (In analogy with spin systems, $x_i$ may be thought of as a single site magnetization and $X$ as the total system magnetization, with $\langle \Delta x^2 \rangle$, $\langle \Delta X^2 \rangle$, $r$ and $R$ the corresponding fluctuations and susceptibilities.)

The fluctuation-dissipation theorem relates these in equilibrium by $\langle \Delta x^2 \rangle = r \cdot T$ and $\langle \Delta X^2 \rangle = R \cdot T$, and inspires the definition in non-equilibrium systems of effective fluctuation temperatures $T_{F}^{(1)} \equiv \langle \Delta x^2 \rangle / r$ and $T_{F}^{(N)} \equiv \langle \Delta X^2 \rangle / R$ for single-particle and many-particle measurements, respectively. In this section we calculate $T_{F}^{(1)}$ and $T_{F}^{(N)}$ for our model and demonstrate that they generally differ one from each other and from both $T_G$ and $T_B$. As in \cite{srebro_levine} we concentrate here on space- and time- independent fluctuation-dissipation relations; their temporal dependence has recently been investigated in \cite{shokef_bunin_levine} and the spatial dependence in \cite{levanony_levine}.

By averaging Eq. (\ref{eq:dyn_rule_fd}) and taking the steady state solution we see that 
\bea
\langle x_i \rangle = \frac{\langle X \rangle}{N} = - \frac{\langle e \rangle} {F_i} , \label{eq:xi_Tg}
\eea
with $\langle e \rangle = T_B f (1-h)/(A_1-fh)$ [this reduces to Eq. (\ref{eq:tg_steady}) in the $h \rightarrow 0$ limit]. Therefore, 
\bea
r = \frac{R}{N} = \frac{\langle e \rangle}{F^2} .
\eea
For the steady state averaged second moments we obtain,
\begin{subequations}
\bea
\langle \Delta x^2 \rangle = \frac{\langle e^2 \rangle}{2F^2} = b(\alpha,f,h,N) \cdot r \cdot T_G , \label{eq:dx}
\eea
\bea
\langle \Delta X^2 \rangle &=& N \langle \Delta x^2 \rangle - \frac{\langle XE \rangle - \langle X \rangle \langle E \rangle}{F} \non \\ &=& B(\alpha,f,h,N) \cdot R \cdot T_G , \label{eq:dX}
\eea
\end{subequations}
where $E \equiv \sum_{i=1}^N e_i$, and $b$ and $B$ are dimensionless functions of the dimensionless model parameters and of the system size. We are interested in the thermodynamic limit ($N \gg 1$), for which $b$ and $B$ reduce to the expressions given in \cite{foot_note_2nd_moms}.

$r$, $R$, $\langle \Delta x^2 \rangle$ and $\langle \Delta X \rangle$ all diverge as $F \rightarrow 0$ (where fluctuation-dissipation relations are normally measured), however their ratios define effective temperatures 
\begin{subequations} \label{eq:tfds}
\bea
T_{F}^{(1)} \equiv \frac{\langle \Delta x^2 \rangle}{r} = b(\alpha,f,h) T_G, \label{eq:tfd1}
\eea
\bea
T_{F}^{(N)} \equiv \frac{\langle \Delta X^2 \rangle}{R} = B(\alpha,f,h) T_G, \label{eq:tfdN}
\eea
\end{subequations}
which are finite and independent of $F$. These fluctuation temperatures generally differ one from the other, are larger than the granular temperature $T_G \equiv \langle e \rangle$ and smaller than the bath temperature $T_B$. Only in the equilibrium limits ($\alpha=1$ and $f=1$) do all effective temperatures coincide with $T_B$. 

\begin{figure}[b]
\includegraphics[width=7cm]{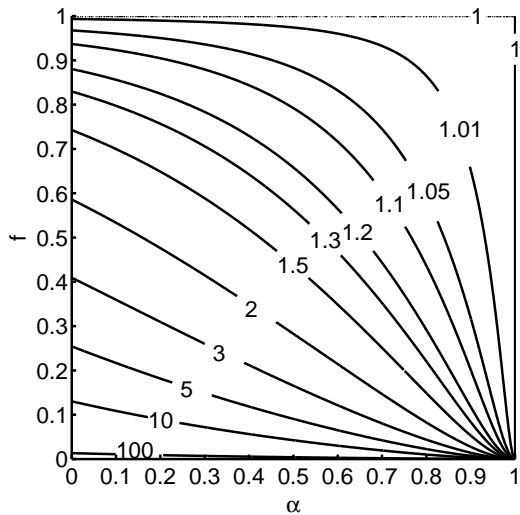}
\caption{\label{fig:tfd_contours_h0} Contours of the ratio $T_{F}/T_G$ between the fluctuation temperatures and the granular temperature vs. the restitution coefficient $\alpha$ and the coupling strength $f$ in the limit of weak coupling between the probe and the system ($h \rightarrow 0$), as given in \cite{foot_note_tfd_h0}. In this limit, $T_{F}^{(N)}=T_{F}^{(1)}$.}
\end{figure}

\begin{figure}[h]
\includegraphics[width=7cm]{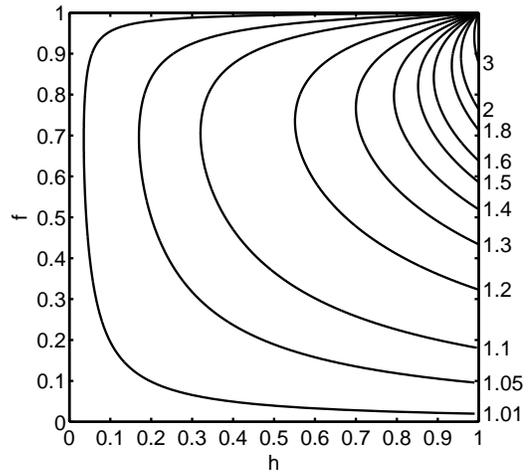}
\caption{\label{fig:tfd_contours_calc_num} Contours of the ratio $T_{F}^{(N)}/T_{F}^{(1)}$ between the many-particle and single-particle fluctuation temperatures vs. the coupling strengths $f$ and $h$ in the maximal dissipation limit ($\alpha = 0$), as given in \cite{foot_note_tfd_a0}.}
\end{figure}

$T_{F}^{(N)}$ is generally larger than $T_{F}^{(1)}$, but in the limit studied in \cite{srebro_levine} of weak coupling between the system and the probe ($h \rightarrow 0$) the two coincide, and the expression for them simplifies to that given in \cite{foot_note_tfd_h0}. Nonetheless, they differ from the granular temperature, as shown in Fig. \ref{fig:tfd_contours_h0}. The difference between the two fluctuation temperatures is most prominent in the maximal dissipation limit ($\alpha = 0$), where they reduce to the expressions given in \cite{foot_note_tfd_a0}. The ratio between them for this case is shown in Fig. \ref{fig:tfd_contours_calc_num}.

The single-particle fluctuation temperature $T_{F}^{(1)}$ directly probes the second moment of the single-particle energy distribution and thus gives the effective temperature $T_R^{(2)}$ defined in Sec. \ref{sec:teffs} [compare Eqs. (\ref{eq:xi_Tg}-\ref{eq:tfds}) to Eq. (\ref{eq:def_tr})]. The many-particle fluctuation temperature $T_{F}^{(N)}$, on the other hand, is defined by a measurement on the entire system, thus reflects correlations between particle energies and cannot be related directly to the effective temperatures defined from the single-particle energy distribution.

In dissipative systems with strong coupling to the driving mechanism ($f \approx 1$) and large restitution coefficient ($\alpha \approx 1$) the energy distribution is close to exponential, and correlations are weak, hence the values of $T_{F}^{(N)}$, $T_{F}^{(1)}$ and $T_G$ are similar (but not identical). As has been predicted by kinetic theory \cite{garzo_2004}, we expect the fluctuation temperatures to be larger than the granular temperature (see Fig. \ref{fig:tfd_contours_h0}) in granular gases as well, where the energy distribution is non-exponential. In the cases studied numerically \cite{puglisi_2002_fd,barrat_2002_fd} the energy distributions were only slightly non-exponential, resulting in small differences between the effective temperatures, which may explain their seeming coincidence.

\subsection{Entropic Temperature}
\label{sec:ts}

In analogy to equilibrium statistical mechanics, a further definition of an effective temperature in non-equilibrium systems may be constructed by differentiating the system's entropy $S$ with respect to its average total energy $\langle E \rangle$, yielding the entropic temperature $T_S \equiv (\partial S / \partial \langle E \rangle)^{-1}$ \cite{barrat_2000,coniglio_2001,makse_2002,t_fd_foam}. It is intriguing to inquire as to how this quantity relates to other effective temperatures of the system and whether it has a fundamental thermodynamic-like significance, stemming from a maximization of entropy upon contact between systems. 

In this section we calculate $T_S$ for the continuous energy model defined in Sec. \ref{sec:definition}, as well as for a discrete energy version (described below). We first identify simple scaling arguments leading to the coincidence of $T_S$ with the granular temperature $T_G \equiv \langle e \rangle$. This scaling holds for the continuous energy model with $\alpha>0$. We then demonstrate the breakdown of this scaling both in the singular limit of the continuous energy model at $\alpha=0$, which exhibits a condensation at $e=0$, and by introducing a discrete energy version of the model, where the structure of the particles' energetic levels contains an additional energy scale $\epsilon$.

\subsubsection{Scaling}

For systems coupled to a heat bath of temperature $T_B$ with no internal energy scale characterizing the system's structure (such as an interaction energy or energy spacing between possible states) the only energy scale in the system is $T_B$: When $T_B$ is varied all energies in the system change linearly with its change. This is the case for the model defined in Sec. \ref{sec:definition}. Different effective temperatures, or energy characteristics of the system, may differ, however they all scale linearly with $T_B$. The ratios between effective temperatures are dimensionless numbers depending on the dimensionless model parameters.

Since the $N$-particle energy distribution has dimension of inverse energy to the $N$'th power, as long as it is non-singular it scales as
\bea 
P(e_1,\ldots,e_N) = \frac{\varphi \left( \frac{e_1}{T_B},\ldots,\frac{e_N}{T_B};\alpha,f \right)}{T_B^N} ,  
\eea
with $\varphi$ a dimensionless function of the dimensionless energies $\{ e_i/T_B \}$ and the dimensionless model parameters ($\alpha$ and $f$ in our case). 

We define the system's entropy as
\bea
S \equiv - \int P(e_1,\ldots,e_N) \ln {P(e_1,\ldots,e_N)} de_1 \cdots de_N \label{eq:S}.
\eea
After changing the integration variables from $\{ e_i \}$ to $\{ e_i/T_B \}$, and using the normalization of $P(e_1,\ldots,e_N)$, we see that 
\bea
S = N \ln{T_B} + const., 
\eea
with the additive constant depending only on the dimensionless parameters $\alpha$ and $f$, but not on $T_B$. Since the energy of the system $\langle E \rangle$ scales linearly with $T_B$ [see Eq. (\ref{eq:tg_steady})], we may write 
\bea \label{eq:SNE}
S = N \ln{\langle E \rangle} + const., 
\eea
and conclude that
\bea
T_S \equiv \left( \frac{\partial S}{\partial \langle E \rangle} \right)^{-1} = \frac{\langle E \rangle}{N} \equiv T_G. 
\eea
The functional form of the dimensionless distribution $\varphi(e_1/T_B,\ldots,e_N/T_B)$, which depends on the model parameters, manifests itself only in the additive term in the entropy, and does not affect the relation $T_S=T_G$.

In order to observe richer behavior we turn to models with additional energy scales. We shall demonstrate this using a definition of $T_S$ which is simpler to calculate, and which for the single energy scale case exactly coincides with the calculation given above. Instead of defining the entire system's entropy $S$ from the entire system's energy distribution [Eq. (\ref{eq:S})], we consider a subsystem comprised of a single particle and measure its entropic temperature. From the single-particle energy distribution $p(e)$, we define the single-particle entropy as
\bea
s \equiv - \int p(e) \ln {p(e)} de \label{eq:s},
\eea
and differentiate it with respect to the particle's average energy: $T_S \equiv (\partial s / \partial \langle e \rangle)^{-1}$. Since all particles have the same single-particle energy distribution, $T_S$ is clearly equal for all particles in the system. The aforementioned scaling argument holds for the single-particle distribution, thus for systems with a single energy scale the single-particle definition gives $T_S=T_G$ as well.

\subsubsection{Condensation in the Maximal Dissipation Model}

For the maximally dissipative limit ($\alpha=0$) of our model, every particle undergoing an interaction with another particle in the system is left with zero energy after the interaction, and the system exhibits a condensation at $e=0$. Since in the steady state for $1-f$ of the particles the last interaction was such an energy draining interaction and not an interaction with the bath, a fraction $1-f$ of the particles have zero energy, and the single-particle energy distribution has the general form
\bea
p(e) = (1-f) \delta(e) + f \frac{ \varphi \left( \frac{e}{T_B} , f \right) } {T_B}. 
\eea

Due to the normalization of $p(e)$, this singular distribution yields an entropy of the form 
\bea 
s = - (1-f) \ln (1-f) + f \ln T_B + const.. 
\eea
The average energy per particle $\langle e \rangle$ scales linearly with $T_B$. Therefore, 
\bea
s = f \ln \langle e \rangle + const., 
\eea
and consequently $T_S = T_G / f$, which is larger than $T_G$. 

\subsubsection{Discrete Energy Model}

Another way to break the scaling arguments leading to $T_S=T_G$ is by introducing an additional energy scale to the model. We now consider every particle as a quantum harmonic oscillator with possible energies $e_i = 0 , \epsilon , 2 \epsilon , ...$. The spacing $\epsilon$ between states constitutes the energy scale which invalidates the scaling arguments presented above. (Even in thermodynamic equilibrium, $T_S=T$ is the equilibrium temperature, while the average energy $\langle e \rangle = \epsilon / [\exp(\epsilon/T)-1]$ differs from the temperature, and $\langle e \rangle \approx T$ only in the continuum limit $\langle e \rangle \gg \epsilon$.) 

The dynamics of the discrete model are as follows: The bath is constructed from similar quantum harmonic oscillators in equilibrium at temperature $T_B$, thus with energies distributed as $p_B(n \epsilon) \sim \exp(-n\epsilon/T_B)$. In the interaction of particle $i$ from the system with a particle of energy $e_B$ from the bath their total energy $e_i + e_B$ is conservatively redistributed between them, by randomly choosing with equal probability a new energy $e_i' \in \{0 , \epsilon , 2 \epsilon , ... , e_i + e_B \}$. In a dissipative interaction between particles $i$ and $j$ of the system, each energy ``quantum'' $\epsilon$ of the total energy $e_T = e_i + e_j$ has a probability $\alpha$ to remain with the interacting pair and a probability $1-\alpha$ to be dissipated out of the system. The remaining energy $e_T'$ is then randomly redistributed between the two particles with equal probability for every outcome $e_i' \in \{0 , \epsilon , 2 \epsilon , ... , e_T' \}$, and $e_j' = e_T' - e_i'$ \cite{footnote_distinguishable}.

\begin{figure}[b]
\includegraphics[width=8cm]{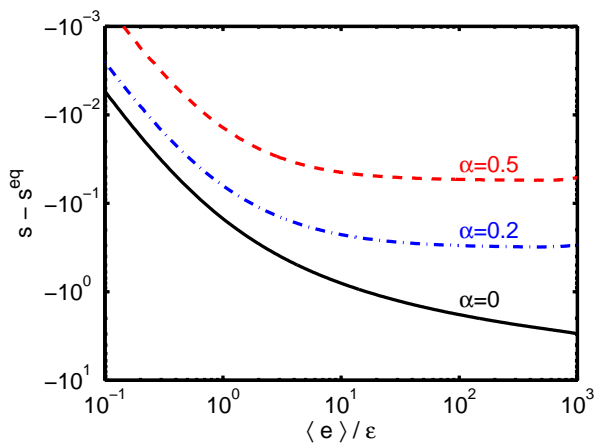}
\caption{\label{fig:tg_ds} (Color online) The deviation of the entropy $s$ from its equilibrium value $s^{eq}$ vs. the average energy $\langle e \rangle$ for several values of the restitution coefficient $\alpha$ and for coupling strength $f=0.5$.}
\end{figure}

We numerically solved this discrete energy model by Monte-Carlo simulation and obtained the average energy as well as the energy distribution $p_n \equiv p(n \epsilon)$, from which the entropy $s \equiv -\sum_{n=0}^{\infty} p_n \ln p_n$ was calculated. We kept the restitution coefficient $\alpha$, the coupling strength $f$, and the energy spacing $\epsilon$ fixed and scanned the bath temperature $T_B$ in order to obtain the dependence of entropy on the average energy for given $\alpha$, $f$ and $\epsilon$. We compare this functional behavior to the corresponding equilibrium behavior (i.e., conservative interactions), where the entropy and average energy of a quantum harmonic oscillator are related by
\bea
s^{eq}(\langle e \rangle) = \ln \left( \frac{\langle e \rangle}{\epsilon} + 1 \right) + \frac{\langle e \rangle}{\epsilon} \ln \left( \frac{\epsilon}{\langle e \rangle} + 1 \right).
\eea
We subtracted $s^{eq}(\langle e \rangle)$ from the numerically obtained $s(\langle e \rangle)$ to yield the deviation from equilibrium behavior of dissipative systems displayed in Fig. \ref{fig:tg_ds}. 

For $\langle e \rangle \gg \epsilon$ the discrete energy model coincides with the continuous energy one: For $\alpha > 0$ the entropy of the dissipative system in this region can be seen to merely be smaller by an additive constant from the entropy of an equilibrium system with the same energy, as expected from the scaling arguments for a system with a single energy scale [Eq. (\ref{eq:SNE})]. For $\alpha=0$, on the other hand, the entropy deviation grows with energy, and $T_S$ differs from $T_G$ even in this high energy limit. 

\begin{figure}[b]
\includegraphics[width=8cm]{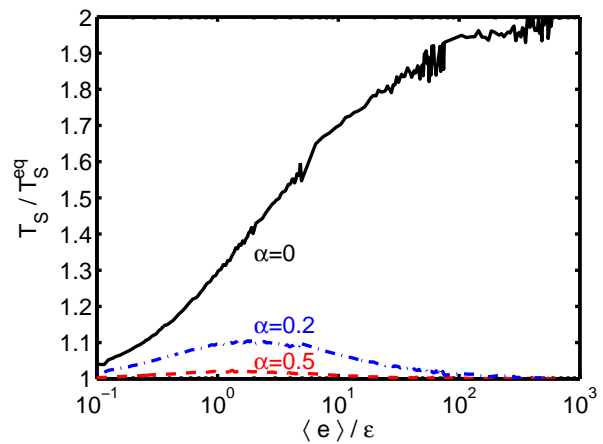}
\caption{\label{fig:ts_tg} (Color online) The ratio $T_S/T_S^{eq}$ of the entropic temperature to its equilibrium value vs. the average energy $\langle e \rangle$ for several values of the restitution coefficient $\alpha$ and for coupling strength $f=0.5$.}
\end{figure}

We numerically differentiated $s(\langle e \rangle)$ with respect to $\langle e \rangle$ to obtain $T_S$, and in Fig. \ref{fig:ts_tg} compare the functional dependence of $T_S$ on $\langle e \rangle$ to the corresponding equilibrium behavior. That is, for every value of $\langle e \rangle$ we normalize $T_S$ by the temperature $T_S^{eq}$ required to give this average energy, were the system in equilibrium. Since the entropy of a dissipative system is smaller than that of an equilibrium system with the same average energy, $T_S(\langle e \rangle)$ is generally larger than $T_S^{eq}(\langle e \rangle)$. For $\alpha>0$, $T_S$ behaves as in equilibrium in the two extremes of very high and very low energy, and exhibits non-equilibrium behavior only for intermediate energies ($\langle e \rangle \approx \epsilon$). In the high energy limit ($\langle e \rangle \gg \epsilon$), the discrete energy model is equivalent to the continuous energy one, thus $T_S$ behaves as in equilibrium for $\alpha>0$, and reaches a value larger by a factor $1/f$ from the equilibrium value for the singular limit $\alpha=0$. 

For very low energies the system behaves as a two level system, irrespective of whether it is in equilibrium or not. For the system to have such a low average energy almost all particles must be in the ground state ($e=0$), and since the occupation of states rapidly decays with energy, only the first excited state ($e = \epsilon$) is relevant, while states of higher energy have a negligible occupation. The energy distribution in a two level system is characterized by a single number (the ratio of occupation of the two states) and is hence not rich enough to exhibit any features of a non-exponential energy distribution. For intermediate energies the entropic temperature exhibits significant deviations from equilibrium behavior even for $\alpha>0$, as clearly seen in Fig. \ref{fig:ts_tg}.

\subsection{Configurational Temperature}
\label{sec:tconf}

The hierarchy of effective temperatures given by $T_R^{(n)}$ [Eq. (\ref{eq:def_tr})] may be related to the hierarchy of the so-called {\it hyperconfigurational temperatures}. A recent extension of the virial theorem, states that for a system in thermodynamic equilibrium at temperature $T$ with dynamics stemming from a Hamiltonian $\mathcal{H}(\{p_i\},\{q_i\})$ ($1 \leq i \leq N$) the following relation holds \cite{jepps_2000}:
\bea
\frac {\langle \vec{\nabla} \mathcal{H} \cdot \vec{B} \rangle}{\langle \vec{\nabla} \cdot \vec{B} \rangle} = T,
\eea
where $\vec{\nabla} \equiv ( \partial / \partial p_1 , ... , \partial / \partial p_N , \partial / \partial q_1 , ... , \partial / \partial q_N )$ represents differentiation with respect to all phase space coordinates, and $\vec{B}$ is an arbitrary vector field in phase space satisfying $0 < | \langle \vec{\nabla} \mathcal{H} \cdot \vec{B} \rangle | < \infty$, $0 < | \langle \vec{\nabla} \cdot \vec{B} \rangle | < \infty$, with $\langle \vec{\nabla} \mathcal{H} \cdot \vec{B} \rangle$ growing slower than $e^N$.

If the Hamiltonian is of the form $\mathcal{H} = \sum_{i=1}^N p_i^2 + V(\{q_i\})$, that is, comprised of a kinetic term depending only on momenta and a potential term depending only on coordinates, it is useful to take $\vec{B} = \vec{\nabla} V$. This yields a relation between the temperature and ensemble averages of solely the particle locations, without the need of measuring momenta. This may be used to define a configurational effective temperature in non-equilibrium systems,
\bea
T_C \equiv \frac {\langle | \vec{\nabla} V |^2 \rangle}{\langle \nabla^2 V  \rangle}.
\eea
The hierarchy of hyperconfigurational temperatures $T_C^{(n)}$ \cite{han_grier_2004} generalizes this by taking $\vec{B} = ( 0 , ... , 0 , (\partial V / \partial q_1)^n , ... , (\partial V / \partial q_N)^n )$, so that
\bea
T_C^{(n)} \equiv \frac {\left\langle \sum_{i=1}^N \left( \frac{\partial V}{\partial q_i} \right)^{n+1} \right\rangle}{\left\langle n \sum_{i=1}^N \left( \frac{\partial V}{\partial q_i} \right)^{n-1} \frac{\partial^2 V}{\partial q_i^2} \right\rangle} \label{eq:t_conf_def}.
\eea

In order to interpret $T_C^{(n)}$ for our case we note that the model's dynamics in the non-dissipative case ($\alpha=1$) manifest a uniform single-particle density of states \cite{srebro_levine_comment}. That is, in the equilibrium limit our system is equivalent to a collection of weakly interacting harmonic oscillators, with $e_i=p_i^2/2+q_i^2/2$, where $\{ q_i \}$ and $\{ p_i \}$ are some hidden coordinates and momenta. Each $q_i$ and $p_i$ change periodically with time as for an isolated harmonic oscillator with energy $e_i$. Occasionally (that is, at a frequency much smaller than the oscillator's frequency) this energy is changed due to an interaction with some other particle or with the bath. It is natural to extend this description to dissipative ($\alpha<1$) cases as well, thus we consider the case where $V(\{ q_i \}) = \sum_{i=1}^N q_i^2/2$. 

For a harmonic oscillator of given energy $e$, the temporal average of $q^{2n}$ over many periods of oscillation is $e^n (2n-1)!! / n!$. The particle's energy changes with time due to interactions, thus we average over its steady state distribution, and obtain $\langle q^{2n} \rangle = \langle e^{n} \rangle (2n-1)!! / n!$. Upon substitution in Eq. (\ref{eq:t_conf_def}) we see that the hyperconfigurational temperatures probe ratios between succeeding moments of the energy distribution, and may thus be related to $T_R^{(n)}$ [defined in Eq. (\ref{eq:def_tr_tm})]:
\bea
T_C^{(2n-1)} \equiv \frac{\langle q^{2n} \rangle}{(2n-1)\langle q^{2n-2} \rangle} = \frac{\langle e^n \rangle}{n \langle e^{n-1} \rangle} \equiv T_R^{(n)}.
\eea
$T_C^{(2n)}$ is undefined since averages of odd moments of $q$ vanish. 

This demonstrates for one particular modeling the connection between the configurational temperature $T_C$, and the single particle energy distribution. We suggest that similar relations hold in other systems, and that the hyperconfigurational temperatures may be used to characterize the non-equilibrium nature of energy distributions in general.

\section{Discussion}
\label{sec:discussion}

This Paper investigates several non-equilibrium phenomena observed in a minimal stochastic model for driven dissipative dynamics. Our model is inspired by granular gases, nevertheless we believe it may be relevant to a broader class of driven dissipative systems. The model is simple enough to admit an exact solution of the single-particle distribution in terms of its moments, in the steady state as well as during the evolution from any initial condition to this state. When considering particles with energies slightly larger than the average energy in the system, the high energy tail of the single-particle distribution is seemingly overpopulated, as has been found in granular gases. However, we have calculated the very high energy tail and found that it behaves exponentially with a decay rate corresponding to the temperature of the driving bath $T_B$. Generally, very high energy tails manifest the bath distribution since dissipative interactions within the system (where typical energies are smaller than in the bath) hardly affect this tail. It will be interesting to investigate the tail of very high energies in other dissipative system, and to test whether they agree with the bath distributions (as has been found in \cite{biben_2002})

Due to the non-exponential energy distributions of non-equilibrium systems, the average energy, or granular temperature $T_G$, is just one energy scale characterizing the system. Higher moments of the distribution define hierarchies of effective temperatures $T_R^{(n)}$ and $T_M^{(n)}$, which generally vary with the order $n$ of the moments, and coincide with the actual temperature if the system is in equilibrium. The large $n$ limit of these effective temperatures relates to the high energy tail of the distribution. In our model these effective temperatures converge to the bath temperature in this limit, reflecting the tail's exponential behavior. Furthermore, we related these effective temperatures to the hyperconfigurational temperatures $T_C^{(n)}$, defined in Hamiltonian system from particle locations. 

Fluctuation-dissipation measurements characterize aspects of the second moment of the energy distribution, and thus show an effective temperature generally differing from the granular temperature (or first moment). Here we have extended the calculations presented in \cite{srebro_levine} to general coupling strength $h$ between the system and the probe, and moreover showed that single-particle and many-particle measurements yield different effective temperatures $T_{F}^{(1)}$ and $T_{F}^{(N)}$. These results, together with those dealing with time dependent measurements presented in \cite{shokef_bunin_levine}, call for examining these phenomena in more realistic dissipative systems.

Finally, we considered the entropic temperature $T_S$ obtained by differentiating the entropy with respect to energy. We showed that generally for a non-singular system without an internal energy scale, simple scaling arguments lead to the exact coincidence of the entropic temperature with the granular temperature. When this scaling is not valid, $T_S$ is generally larger than the corresponding equilibrium value. It is intriguing to test this scaling and its breakdown in the more complex systems where $T_S$ can been measured.

We can identify an ordering of effective temperatures in our model: For the hierarchies defined from the energy moments, both $T_R^{(n)}$ and $T_M^{(n)}$ grow with $n$; The fluctuation temperatures $T_{F}$ are larger than the granular temperature $T_G$; The many-particle fluctuation temperature $T_{F}^{(N)}$ is larger than the single-particle one $T_{F}^{(1)}$; All these effective temperatures are smaller than the bath temperature $T_B$. It will be interesting to see whether such ordering occurs in other driven dissipative systems as well. 


\begin{acknowledgments}

We thank Naama Brenner, Guy Bunin, Bernard Derrida, J. Robert Dorfman, Sam F. Edwards, Dmitri Grinev, Fred MacKintosh, Dana Levanony, Yael Roichman and Gal Shulkind for helpful discussions. D.L. acknowledges support from Grants No. 88/02 and No. 660/05 of the Israel Science Foundation and the Fund for the Promotion of Research at the Technion.

\end{acknowledgments}




\begin{thebibliography}{10}

\bibitem{noije_1998}
T.P.C. van Noije and M.H. Ernst, Granular Metter {\bf 1}, 57 (1998).

\bibitem{bennaim_krapivsky_2000}
E. Ben-Naim and P.L. Krapivsky, Phys. Rev. E {\bf 61}, R5 (2000).

\bibitem{moon_2001}
S.J. Moon, M.D. Shattuck, and J.B. Swift, Phys. Rev. E {\bf 64}, 031303 (2001).

\bibitem{moon_2004}
S.J. Moon, J.B. Swift, and H.L. Swinney, Phys. Rev. E {\bf 69}, 011301 (2004).

\bibitem{zon_2004}
J.S. van Zon and F.C. MacKintosh, Phys. Rev. Lett. {\bf 93}, 038001 (2004).

\bibitem{zon_2005}
J.S. van Zon and F.C. MacKintosh, Phys. Rev. E {\bf 72}, 051301 (2005).

\bibitem{puglisi_2002_fd}
A. Puglisi, A. Baldassarri, and V. Loreto, Phys. Rev. E {\bf 66}, 061305 (2002).

\bibitem{barrat_2002_fd}
A. Barrat, V. Loreto, and A. Puglisi, Physica A {\bf 334}, 513 (2004).

\bibitem{garzo_2004}
V. Garz$\acute{o}$, Physica A {\bf 343} 105 (2004).

\bibitem{langer_2000}
S.A. Langer and A.J. Liu, Europhys. Lett. {\bf 49}, 68 (2000).

\bibitem{t_fd_foam}
I.K. Ono, C.S. O'Hern, D.J. Durian, S.A. Langer, A.J. Liu, and S.R. Nagel, Phys. Rev. Lett. {\bf 89}, 095703 (2002).

\bibitem{ohern_2004}
C.S. O'Hern, A.J. Liu, and S.R. Nagel, Phys. Rev. Lett. {\bf 93}, 165702 (2004).

\bibitem{barrat_2000}
A. Barrat, J. Kurchan, V. Loreto, and M. Sellitto, Phys. Rev. Lett. {\bf 85}, 5034 (2000).

\bibitem{barrat_2001}
A. Barrat, J. Kurchan, V. Loreto, and M. Sellitto, Phys. Rev. E {\bf 63}, 051301 (2001).

\bibitem{berthier_2002}
L. Berthier and J.L. Barrat, Phys. Rev. Lett. {\bf 89},  095702 (2002).

\bibitem{biben_2002}
T. Biben, P.A. Martin, and J. Piasecki, Physica A {\bf 310}, 308 (2002).

\bibitem{coniglio_2001}
A. Coniglio and M. Nicodemi, Physica A {\bf 296}, 451 (2001).

\bibitem{makse_2002}
H.A. Makse and J. Kurchan, Nature {\bf 415}, 614 (2002).

\bibitem{tg}
P.K. Haff, J. Fluid Mech. {\bf 134}, 401 (1983).

\bibitem{hohenberg_1989}
P.C. Hohenberg and B.I. Shraiman, Physica D {\bf 37}, 109 (1989).

\bibitem{cugliandolo_1997}
L.F. Cugliandolo, J. Kurchan, and L. Peliti, Phys. Rev. E {\bf 55}, 3898 (1997).

\bibitem{rugh_1997}
H.H. Rugh, Phys. Rev. Lett. {\bf 78}, 772 (1997).

\bibitem{jepps_2000}
O.G. Jepps, G. Ayton, and D.J. Evans, Phys. Rev. E {\bf 62}, 4757 (2000).

\bibitem{han_grier_2004}
Y. Han and D.G. Grier, Phys. Rev. Lett. {\bf 92}, 148301 (2004).

\bibitem{powels_2005}
J.G. Powels, G. Rickayzen, and D.M. Heyes, Molecular Physics {\bf 103}, 1361 (2005).

\bibitem{srebro_levine}
Y. Srebro and D. Levine, Phys. Rev. Lett. {\bf 93}, 240601 (2004).

\bibitem{ulam_1980}
S. Ulam, Adv. Appl. Math. {\bf 1}, 7 (1980).

\bibitem{rouyer_2000}
F. Rouyer and N. Menon, Phys. Rev. Lett. {\bf 85}, 3676 (2000).

\bibitem{feitosa_menon_2002}
K. Feitosa and N. Menon, Phys. Rev. Lett. {\bf 88}, 198301 (2002).

\bibitem{kob_1997}
W. Kob, C. Donati, S.J. Plimpton, P.H. Poole, and S.C. Glotzer, Phys. Rev. Lett. {\bf 79}, 2827 (1997).

\bibitem{kasper_1998}
A. Kasper, E. Bartsch, and H. Sillescu, Langmuir {\bf 14}, 5004 (1998).

\bibitem{marcus_1999}
A.H. Marcus, J. Schofield, and S.A. Rice, Phys. Rev. E {\bf 60}, 5725 (1999).

\bibitem{weeks_2000}
E.R. Weeks, J.C. Crocker, A.C. Levitt, A. Schofield, and D.A. Weitz, Science {\bf 287}, 627 (2000)

\bibitem{krook_wu}
M. Krook and T.T. Wu, Phys. Rev. Lett. {\bf 36}, 1107 (1976).

\bibitem{srebro_levine_comment}
Y. Srebro and D. Levine, Phys. Rev. Lett. {\bf 94}, 208901 (2005).

\bibitem{shokef_bunin_levine}
Y. Shokef, G. Bunin, and D. Levine, Phys. Rev. E {\bf 73}, 046132 (2006).

\bibitem{levanony_levine}
D. Levanony and D. Levine, Phys. Rev. E {\bf 73}, 055102(R) (2006).

\bibitem{foot_note_2nd_moms}
For $N \gg 1$ we obtain \\
%
$ b = \{ f(h^2+2h-1) + 2[(1-3h)+A_1/f]A_1 \non \\ + (1-h)(3-A_2) \} / [(1-h)(2A_2-fh)]$, \\
%
$ B = \boldsymbol{(} 
fh \{ [18(1-h) -3f(4-h-7h^2) \non \\
+f^2(2+3h-15h^2-6h^3)]+2[15-27h  \non \\
-f(5-6h-21h^2)]A_1 -[18(1-h) \non \\
+f(3h+1)(3h-2)]A_2 -2(5-9h)A_1A_2 \} \non \\
+2A_1^2 \{ [9-3h -f(3-10h+21h^2)] -(3-h)A_2  \non \\ 
+6/f[A_1 +f(1-2h)]A_1 \} 
\boldsymbol{)} \non \\ 
/[6(1-h)(A_1-fh)(A_1+fh)(2A_2-fh)]$.

\bibitem{foot_note_tfd_h0}
For $h \rightarrow 0$ we obtain \\
%
$b = B = [(4-f)(1-f+2A_1)- \\ (4-3f)A_2]/(2 f A_2)$.
%

\bibitem{foot_note_tfd_a0}
For $\alpha=0$ we obtain \\
%
$ b = \{ 4(2-f)+fh[f(h+7)-12] \} / \\ \{ f(1-h)[6-f(2+h)] \} $ , \\
%
$ B = [ 32-16f(3+2h)+2f^2(12+27h-11h^2) \non\\
-f^3(1-h)(4+38h+13h^2) +f^4h(7+h)(1-h-h^2)] \non\\ 
/  \{ f(1-h) [6-f(2+h)] [(2-f)^2-f^2h^2] \} $. 
%

\bibitem{footnote_distinguishable}
Note that energy ``quanta'' are dissipated as if they were distinguishable entities, whereas for redistributing the remaining energy they are treated as indistinguishable.

\end{thebibliography}
\end{document}